\documentclass[runningheads]{llncs}

\usepackage{booktabs} 
\usepackage{todonotes}
\usepackage{censor}
\usepackage{siunitx}

\begin{document}
\title{Query by Semantic Sketch}

\author{Luca Rossetto\inst{1,2} \and Ralph Gasser\inst{1} \and Heiko Schuldt\inst{1}}
\institute{
Department of Mathematics and Computer Science\\
University of Basel, Basel, Switzerland\\
\email{\{firstname.lastname\}@unibas.ch} \\
\and
Department of Informatics, University of Zurich, Zurich, Switzerland\\
\email{rossetto@ifi.uzh.ch}
}

\maketitle

\begin{abstract}

Sketch-based query formulation is very common in image and video retrieval as these techniques often complement textual retrieval methods that are based on either manual or machine generated annotations. In this paper, we present a retrieval approach that allows to query visual media collections by sketching concept maps, thereby merging sketch-based retrieval with the search for semantic labels.
Users can draw a spatial distribution of different concept labels, such as ``sky'', ``sea'' or ``person'' and then use these sketches to find images or video scenes that exhibit a similar distribution of these concepts. Hence, this approach does not only take the semantic concepts themselves into account, but also their semantic relations as well as their spatial context. The efficient vector representation enables efficient retrieval even in large multimedia collections. We have integrated the \emph{semantic sketch query mode} into our retrieval engine vitrivr and demonstrated its effectiveness.

\end{abstract}


\section{Introduction}

Effective and efficient retrieval of information from large media collections remains a challenge that is still subject to research from many different domains in computer science. As these multimedia collections grow larger, finding a particular item of interest becomes more challenging. This can mainly be attributed to the fact that the traditional and still very common approach of textually annotating individual media objects and retrieving them later based on these labels or descriptions does not scale well -- especially nowadays, as media collections grow at an ever increasing pace. Furthermore, textual descriptions are often subject to bias or error, and would even have to anticipate future interpretations of an object. In addition, in some cases textual descriptions may not be well suited to, for example, describe the temporal changes inherent to a video or the complex scenery that makes up an image.

Content-based retrieval approaches try to remedy this situation by deriving features directly from the media objects' content and to retrieve similar objects w.r.t.\ these features. This removes the necessity for a manual, laborious and tedious annotation process. Common query modes for the visual domain, i.e., videos and images, include Query-by-Example (QbE) or Query-by-Sketch (QbS). These query modes can be summarized as finding an object based on the similarity of that object to either a user-provided example image (QbE) or a user-generated sketch (QbS). For instance, a user could try to use some similar image they found on Google to retrieve the specific image or scene they are looking for. Or they could try to come up with a sketch that resembles the content of the desired object, either as an outline (line sketch), or in terms of colour distribution. Also combinations of QbE and QbS would be possible, for instance by using a query image and superimposing a sketch to add information not present on the query image.

Commonly, the aforementioned query modes take visual similarity based on either color, edge or point-of-interest-based features into account. However, reproducing the visual content of an image by means of a hand-drawn sketch is very difficult in practice and retrieval quality often suffers because of the user's inability to do so.
Inspired by and improving upon the work in~\cite{furuta2018efficient}, we present a video and image retrieval method that exploits concept-maps~\cite{xu2010image} for sketch-based retrieval. Our approach facilitates retrieval from huge image or video collections, such as the V3C~\cite{rossetto2019v3c} used for TRECVID. It allows to search for simple concept labels, such as ``person'', ``sky'' or ``horse'',  while at the same time constraining the occurrence of these labels to specific areas in the image or key-frame. Leveraging recent progress in the area of machine learning and image understanding, we use a deep neural network to produce these concept maps from which we subsequently derive our features for retrieval. Queries can then be expressed by the user, simply by drawing the spatial distribution of the desired concepts as they can recall it from memory. The \emph{semantic sketch query mode} has been implemented into the retrieval engine vitrivr \cite{rossetto2016vitrivr,gasser2019towards} and successfully field-tested on different occasions.

The contribution of this paper is two-fold: First, we introduce a compact vector representation that can be used for efficient sketch-based retrieval of semantic concepts using kNN-lookup, conserving both the spatial as well as the semantic relationship of the sketched objects. Second, we compare our new vector representation to similar methods and we present performance metrics obtained on the V3C1 video collection \cite{rossetto2019v3c} using vitrivr, our multimedia information retrieval stack.

The remainder of this paper is structured as follows: Section~\ref{sec:relatedwork} surveys related work. Section~\ref{sec:methods} describes how we derive the features from an input image and Section~\ref{sec:implementation} details implementation aspects. Section~\ref{sec:evaluation} provides an overview of the method's performance and Section~\ref{sec:conclusion} concludes and offers some outlook to future work.

\section{Related Work}
\label{sec:relatedwork}

The approach of sketch-based retrieval of visual content such as images or videos has been in use for many years as a way to reduce the media discontinuity that occurs when text (or another non-visual modality) is used for query expression. Earlier approaches primarily rely purely on low-level features, such as properties extracted from edge~\cite{hu2010gradient,saavedra2010improved} or color~\cite{bui2015iccv,lokoc2017color,niblack1993qbic} information. Either independently or in conjunction with these sketches, semantic information can be expressed by keywords or descriptions that are derived from manual or machine generated annotations~\cite{densecap,vinyals2017show}. There are some approaches that try to bridge the divide between the visual sketches and the textual annotations, by generating the semantic labels based on sketched input~\cite{eitz2012humans,tanase2016semantic}. However, these methods differ mostly in the user-facing query formulation stage and do not offer any different or even richer query information. In addition, none of these approaches are able to incorporate the spatial relation of the semantic concepts into a query.

With the advent of deep learning-based image and video analysis and its influence on multimedia information retrieval, querying for semantic concepts, while at the same time considering the spatial relation of these concepts within the visual content, became feasible without a textual representation. One way of achieving this goal is by mapping both the query and the target images into a common space that encodes both the visual as well as the semantic and spatial information. Such transformations exist both for image sketches~\cite{gordo2016deep} and for annotated region `sketches', usually consisting of placements of concepts, potentially accompanied by a bounding box delineating the spatial extension of the concept~\cite{mai2017spatial,xu2010image}. Similar approaches have also been used to retrieve results that are both visually and semantically similar to a given sketch~\cite{sangkloy2016sketchy}.

Another technique has been proposed in~\cite{redmon2018yolov3}. The authors use neural networks that generate bounding boxes for the detected instances of semantic concepts. A query, which must also be composed of semantically labelled boxes, can then be answered without the need for a dedicated learned space~\cite{amato2019visione}. Semantic queries based on labelled rectangular boxes have also already been proposed before the advent of deep learning~\cite{lim2001building}, some even taking additional information such as the approximate color of the relevant object into account~\cite{liu2010robust}.

\begin{figure}[t]
\centering
\includegraphics[width=0.85\textwidth]{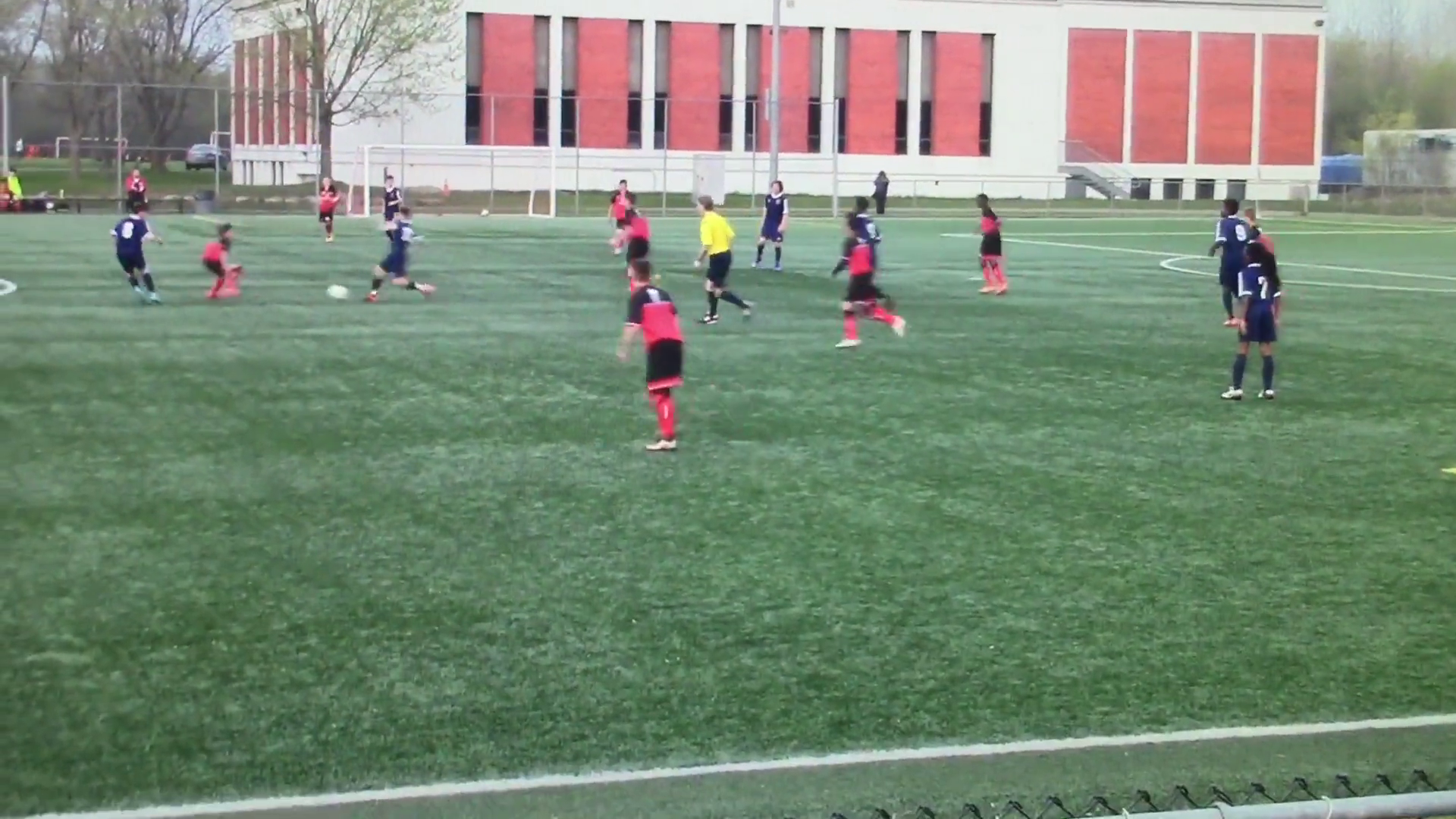}
\caption{Example input image used for the DeepLab pixel-wise annotation. Image cc-by-nc `alpopescu', taken from video 04507 of the V3C~\cite{rossetto2019v3c} video dataset.}
\label{fig:input}
\end{figure}

The approach presented in \cite{furuta2018efficient} goes beyond labelled, rectangular areas by employing a fully convolutional neural network architecture to obtain pixel-wise semantic labels for an input image. The output of such a network can then be linearized and transformed into a binary vector using a one-shot encoding scheme for every pixel. In order to handle the resulting vectors with their dimensionalities ranging up to the order of $10^5$, the authors use product quantization (PQ)~\cite{jegou2011product}. The resulting representation enables true sketch-based retrieval with a high spatial resolution.

Our method builds upon the ideas in \cite{furuta2018efficient} and improves upon their method by introducing a more compact vector representation that does not rely on a specific indexing scheme for efficient retrieval. We achieve this, while simultaneously preserving the semantic relations between the different concepts supported in a query.

\section{Methods}
\label{sec:methods}
\begin{figure}[t]
\centering
\includegraphics[width=0.85\textwidth]{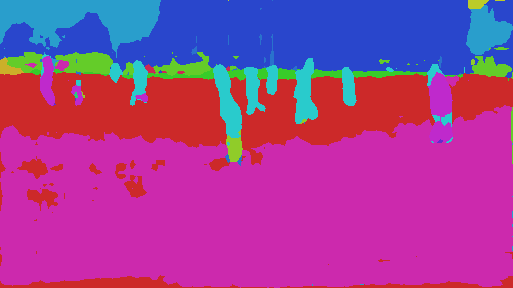}
\caption{Pixel-wise assignment of labels generated by DeepLab based on the input image depicted in Figure \ref{fig:input}.}
\label{fig:assignment}
\end{figure}

In order to transform an input image or video frame into a representation suitable for retrieval via a semantic sketch query, we first apply the DeepLab pixel-wise concept annotation described in \cite{chen2018encoder}, which assigns a concept label to every pixel in the image. This process is exemplified in Figures \ref{fig:input} and \ref{fig:assignment}, which depict an input image and the resulting map corresponding to the assignment of labels.

\begin{figure}
    \centering
    \includegraphics[width=0.6\textwidth]{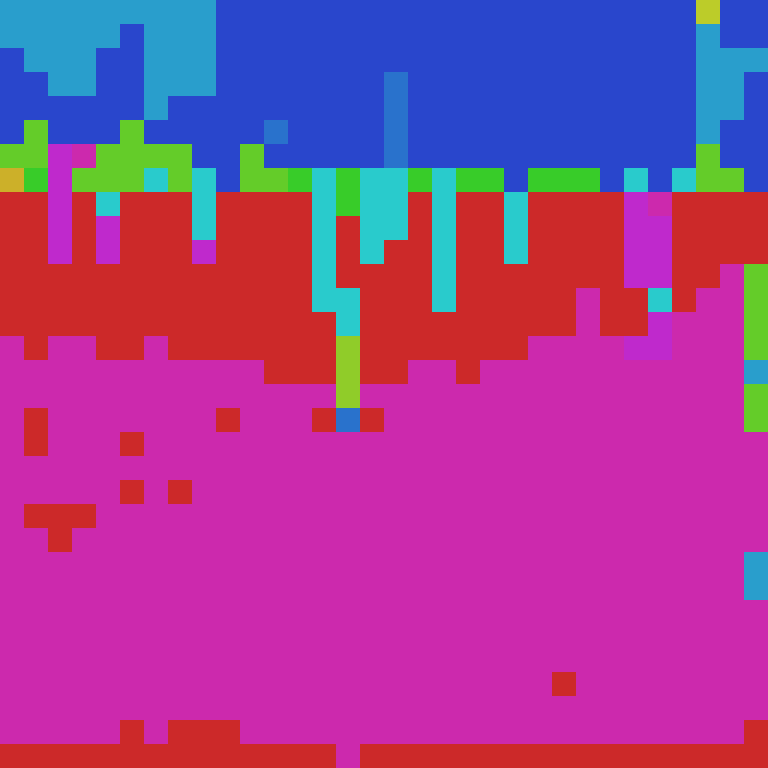}
    \caption{Aggregation of the pixel-wise labels illustrated in Figure~\ref{fig:assignment} into a $32 \times 32$ grid.}
    \label{fig:grid}
\end{figure}

Since the resolution of this pixel-wise label assignment is higher than necessary for a sketch-based application, the map is subsequently down-sampled to an $n \times n$ grid. For each grid cell, all the labels that fall within that cell are used to characterize it. The resulting label of the cell is then simply the label which occurs most often within the cell. This sampling method is chosen for its simplicity, since it does not require any additional information about the semantic relation of the concept labels. In case such relations were available -- e.g., as hierarchical structure between the labels, such as~\cite{miller1995wordnet} -- other methods that preserve more of the semantic information could be employed. Figure~\ref{fig:grid} illustrates the resulting grid representation for a $n \times n$ grid with $n = 32$. This procedure is equivalent to a non-aspect ratio preserving down-scaling of the output map and it hence preserves the spatial relation of the classes as much as the reduced spatial resolution allows.

\begin{figure*}[t]
    \centering
    \includegraphics[width=0.95\textwidth]{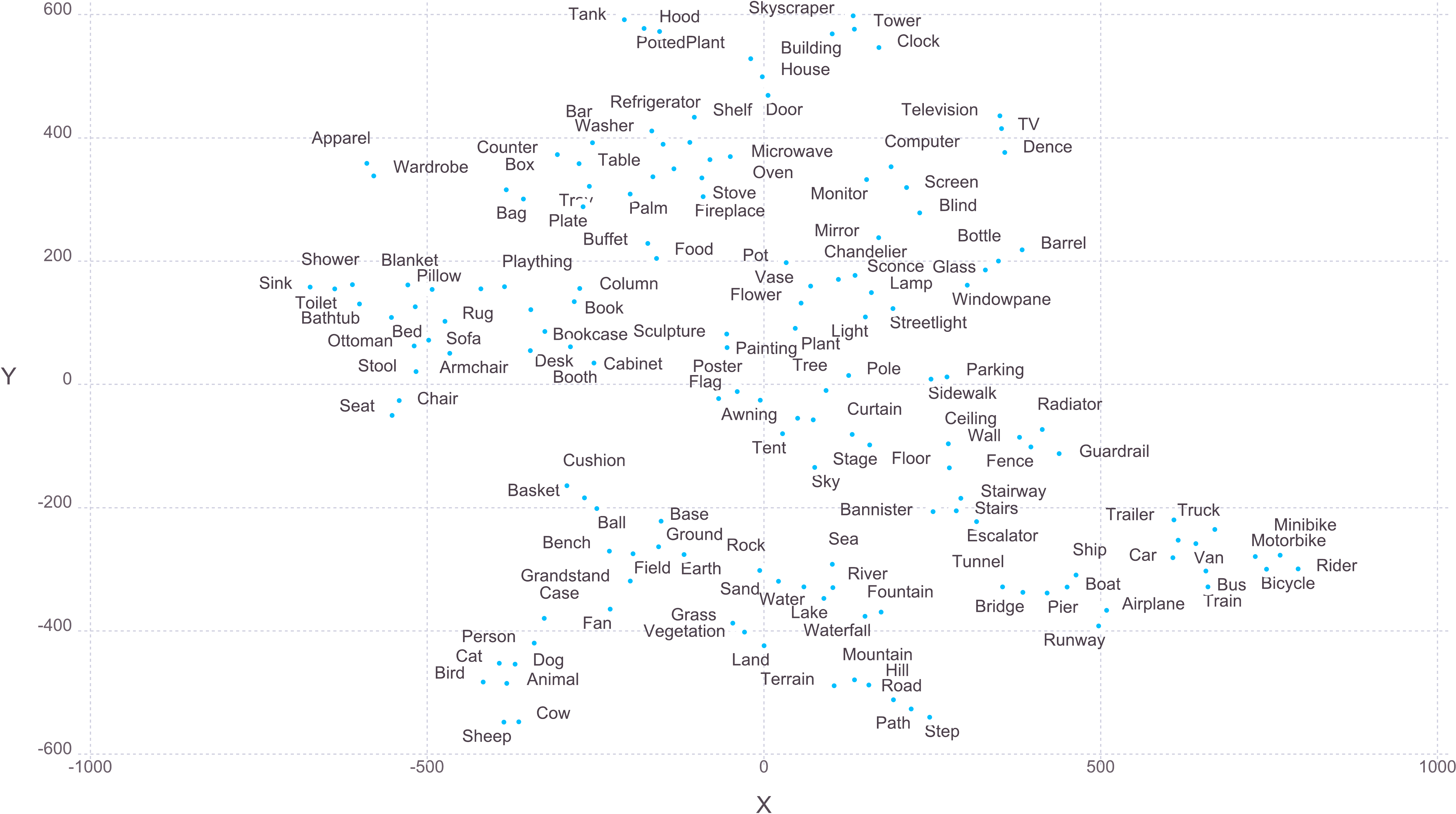}
    \caption{t-SNE embedding of \emph{Word2vec} representations of detectable semantic concepts}
    \label{fig:tsne}
\end{figure*}

The next step involves linearizing this 2D map into a 1D representation, which is required in order to be useful for nearest neighbor retrieval. Rather than just assigning arbitrary scalar values to the individual concepts, we use \emph{Word2vec}~\cite{mikolov2013efficient} to obtain a representation of the concepts, which allows us to reason about their respective semantic similarity. Since the vectors produced for every concept are too long to be used for the map representation directly, we reduce their dimensions to a substantially lower number $d$ (typically 2 or 3) by applying a \emph{t-SNE} embedding~\cite{maaten2008visualizing}. A possible result of such an embedding for $d = 2$ is illustrated in Figure~\ref{fig:tsne}. The resulting coordinates for every label can finally be concatenated by traversing the aggregated two-dimensional label map in a fixed order. The resulting vector has a dimensionality of $n^2d$ (in our case $32^2 \cdot 2 = 2048$). This  is substantially lower than the one produced by \cite{furuta2018efficient}, which produces binary vectors of 245'760 dimensions. Nevertheless, it still retains sufficient information so as to be useful for retrieval.

The query vector is constructed analogously from a pixel-wise semantic annotation that can be sketched by a user. This sketch is transformed to the same grid representation and linearized in the same order, so that the grid cells are aligned between query and feature vectors, thus preserving the spatial relation. 

\begin{figure}[t]
\centering
\includegraphics[width=0.85\textwidth]{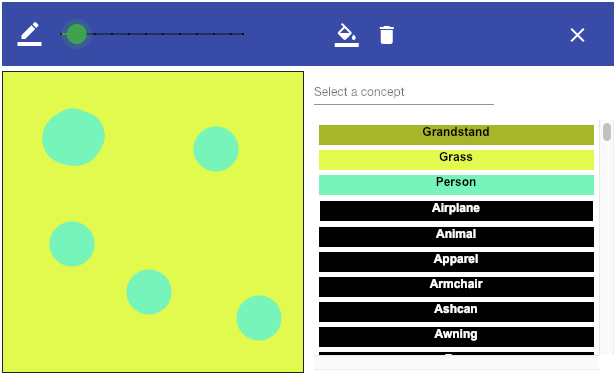}
\caption{Example of a query image as created by a user via the semantic sketch user interface. In this example, the user is employing the concepts ``grass'' and ``person''.}
\label{fig:queryformulation}
\end{figure}

Using this encoding scheme has the advantage that it considers both spatial as well as semantic similarity when comparing two images. For example, a query for a dog lying in the grass will return a lower distance for images showing a cat in the grass than for images of grass without an animal in it, which would not be the case if arbitrary scalar substitutes for the labels had been used. The proposed method also supports non-rectangular areas for one semantic class, which enables more precise representation and querying in comparison to bounding-box based annotations such as those produced by \cite{redmon2018yolov3}.

\section{Implementation}
\label{sec:implementation}
This section describes some details of our implementation, both in terms of feature extraction and retrieval.

\begin{figure*}[t]
\centering
\includegraphics[width=0.95\textwidth]{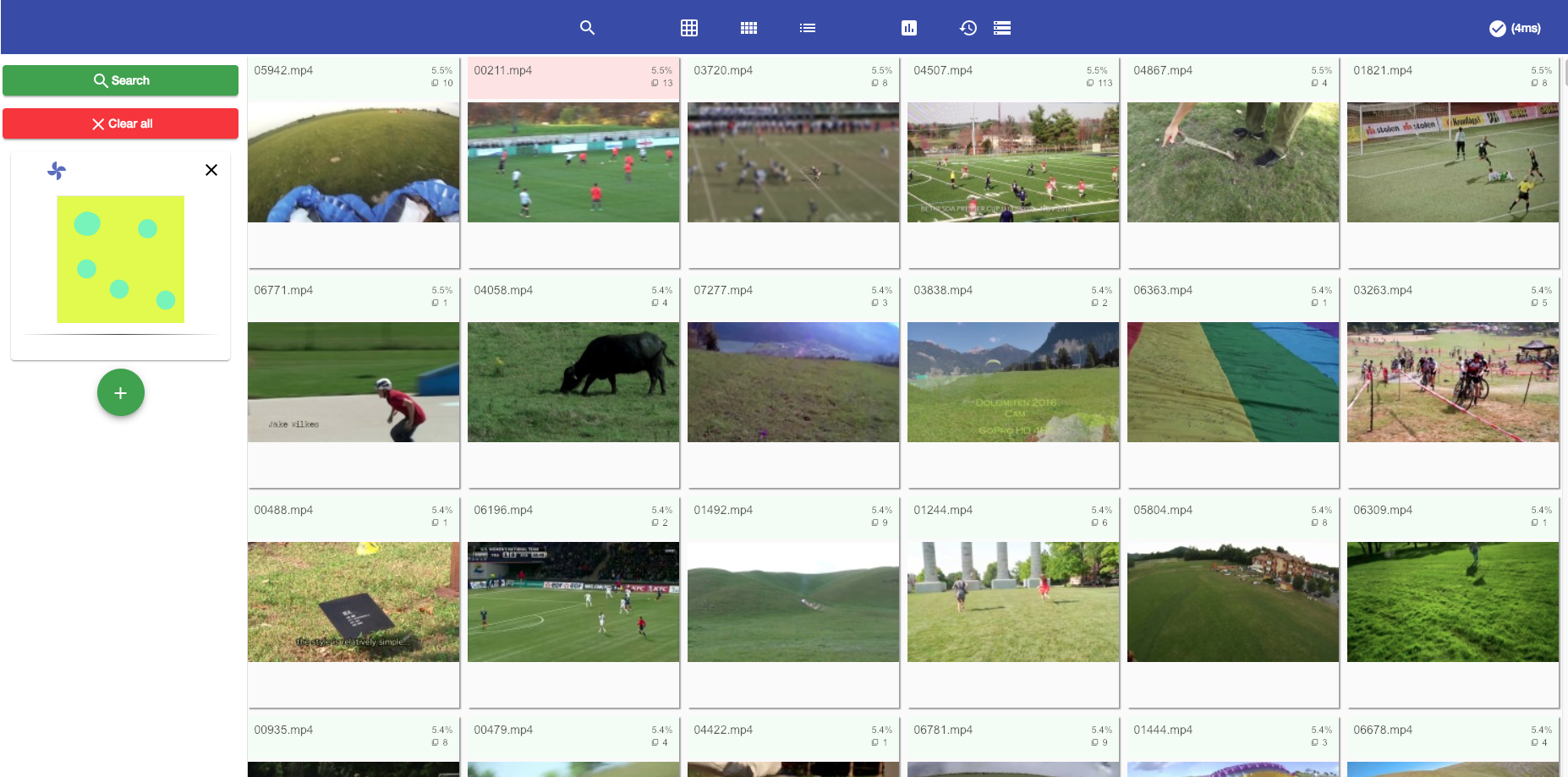}
\caption{Query results returned by the semantic sketch query depicted in Figure \ref{fig:queryformulation}. In this example, we searched for the concepts ``grass'' (yellow) and ``person'' (cyan). The results were produced using the V3C video collection~\cite{rossetto2019v3c}.}
\label{fig:queryresults}
\end{figure*}

\subsection{Extraction}
We use the TensorFlow~\cite{abadi2016tensorflow} framework to run inference on the DeepLab network. Several instances of the same network can be run, each trained on one of the three different datasets such as Cityscapes~\cite{cordts2016cityscapes}, PascalVOC~\cite{everingham2015pascal} and ADE20K~\cite{zhou2017scene}, pre-trained versions of which have been provided by the authors of~\cite{chen2018encoder}. 
We refer to their repository on GitHub\footnote{\url{https://github.com/tensorflow/models/tree/master/research/deeplab}} for more information.

Each of these networks is applied to all visual inputs (either individual images or key frames generated from videos) as part of the extraction pipeline. The outputs of these independent neural networks are then combined during the spatial aggregation step. Since the aggregation simply uses the most common label per cell, it can easily handle labels originating from an arbitrary number of sources without any further requirements towards the sources themselves.
This combination procedure enables us to extend the range of detected semantic concepts without the need for retraining the network on a combined dataset. Since this increases extraction time, network instances can be disabled individually depending on their applicability to the available content. Adding additional semantic classes does therefore not necessitate the retraining of any existing neural network classifiers, the embedding however needs to be updated in such cases.

The \emph{Word2vec} model used for the representation of the semantic concepts is available pre-trained on the \emph{Google News Corpus}\footnote{\url{https://code.google.com/archive/p/word2vec/}}. Since the concepts are known beforehand and remain constant, both the word vectors and their low-dimensional embedding can be pre-computed and do therefore not affect extraction time.

\subsection{Retrieval}

For retrieval, the user is required to sketch the spatial distribution of the different concepts on a 2D canvas. They can pick the concepts from a palette, and each concept is represented by a random, yet distinct color. Therefore, the user manually reproduces the aforementioned pixel-wise 2D assignment of concepts to pixels. This is illustrated in Figure \ref{fig:queryformulation}.

The resulting 2D map plus an assignment of colors to concepts is subsequently sent to the retrieval engine and undergoes the processing described in Section~\ref{sec:methods}, hence yielding a $n^2d$ vector representation of the original sketch. This query vector is then used to perform a kNN-lookup using the Manhattan (L1) metric. The top $k$ results are subsequently returned to the UI and displayed to the user. The results for the query in Figure~\ref{fig:queryformulation} are depicted in Figure \ref{fig:queryresults}. 

\section{Evaluation}
\label{sec:evaluation}
To demonstrate the feasibility of our approach, we evaluate the retrieval time on the V3C1 dataset~\cite{rossetto2019v3c}. For our test collection of $1046235$ vectors, a kNN-lookup leveraging naive, linear scanning of the table took $\SI{974}{ms}$ on average (Intel Core i7@2.6GHz, 32GB RAM; all the data was stored on a Samsung T5 Portable SSD and read from disk). In order to support even larger collections or to further reduce retrieval time, the kNN-lookup can be supported by both precise and approximate index structures. In contrast to \cite{furuta2018efficient}, no particular index is required by our method. The interactive nature of this method has also been demonstrated publicly on multiple occasions~\cite{rossetto2019retrieval,gasser2019multimodal}.

Since our proposed method has several free parameters that can be adjusted to suit a particular use case, it is not possible to make any definite statements about the storage requirements. Table~\ref{tab:storage} therefore shows several possible configurations and compares their storage footprint with those of \cite{furuta2018efficient}, which we use as a baseline. It can be seen that our proposed method can reduce the storage requirements substantially.

\begin{table}[h]
\centering
\caption{Storage requirement comparison of the method described in \cite{furuta2018efficient} with our proposed method using different parameters for \emph{spatial aggregation}, \emph{number of dimensions} for the semantic embedding as well as the \emph{data type}  for storing numbers (bits per dimension).}
\label{tab:storage}
\begin{tabular}{|l|r|r|r|r|}
\hline
Method & Spatial Aggregation & Semantic Dimensions & Bits/Dimension & Storage use \\ \hline
\cite{furuta2018efficient} (baseline)      & N/A                 & N/A                  & 1              & 100\%       \\ \hline
our method  & 32                   & 2                    & 32             & 26.7\%      \\ \hline
our method   & 16                  & 3                    & 32             & 10\%        \\ \hline
our method   & 8                   & 2                    & 8              & 4.2\%       \\ \hline
\end{tabular}
\end{table}

\section{Conclusion and Future Work}
\label{sec:conclusion}

In this paper, we have presented a practical approach for retrieval of visual content based on semantic sketches. Our method produces vector representations that are substantially more compact than those produced by previous techniques and does therefore not necessitate the use of any dedicated index structures for effective retrieval. The length of the produced vectors is dependent on several parameters which can be tuned to adjust the trade-off between retrieval time and spatial as well as semantic accuracy to the relevant use case. How much the query time can be further reduced by the use of such index structures remains subject to future evaluations. The proposed method is also able to take the semantic similarity between the supported classes into account and enables users to easily add support for new semantic classes without the need for retraining of existing neural network classifiers. Future work will investigate how more stable embedding alternatives to t-SNE, such as umap~\cite{mcinnes2018umap} can be used to simplify the process of adding further semantic classes.

Our next step will be to further demonstrate the efficiency and effectiveness of our approach, which will be the scope of a larger study. The exploration of different combination strategies for the results produced by several network instances as well as the aforementioned use of other embedding and dimensionality reduction methods also remains subject to future work.

\bibliographystyle{splncs04}
\bibliography{bibliography}

\end{document}